\documentstyle[12pt]{article}

\begin{document}
\begin{center}
{\large Variable Speed of Light Cosmology and Bimetric Gravity: An
Alternative to Standard Inflation}
\vskip 0.3 true in {\large J. W. Moffat} \vskip 0.3 true in {\large
Perimeter Institute for Theoretical Physics, Waterloo, Ontario,
N2J2W9, Canada}
\vskip 0.3 true in and
\vskip 0.3 true in {\large
Department of Physics, University of Toronto, Toronto, Ontario, M5S1A7, Canada}
\end{center}
\begin{abstract}%
A scalar-tensor bimetric gravity model of early universe cosmology is
reviewed. The metric frame with a variable speed of light (VSL) and a
constant speed of gravitational waves is used to describe a
Friedmann-Robertson-Walker universe. The Friedmann equations are solved
for a radiation dominated equation of state and the power spectrum is
predicted to be scale invariant with a scalar mode spectral index
$n_s=0.97$. The scalar modes are born in a ground state superhorizon and
the fluctuation modes are causally connected by the VSL mechanism. The
cosmological constant is equated to zero and there is no significant
dependence on the scalar field potential energy. A possible way
of distinguishing the bimetric gravity model from standard
inflationary models is discussed. \end{abstract} \vskip 0.2
true in e-mail: jmoffat@perimeterinstitute.ca
\vskip 0.5 true in
Talk given at the Coral Gables Conference on High Energy Physics and
Cosmology, Fort Lauderdale, Florida, December 17-21, 2003.

\section{\bf Bimetric Gravity Theory}

In the scalar-tensor bimetric gravity theory, the metric of
spacetime is given
by~\cite{Clayton,Clayton2,Clayton3,Clayton4}\footnote{Bekenstein~\cite{Bekenstein}
considered bimetric (disformal) gravity theories in the context
of lensing effects in generalized gravity theories, not as
variable speed of light cosmological theories}: \begin{equation}
{\hat g}_{\mu\nu}=g_{\mu\nu}+B\partial_\mu\phi\partial_\nu\phi,
\end{equation} where $B=[\rm length]^2$ and $\phi$ is a scalar
field (biscalar field). The metric ${\hat g}_{\mu\nu}$ is called
the `matter' metric and $g_{\mu\nu}$ is the gravitational metric.
The biscalar field $\phi$ is a component of the gravitational
field.

The model consists of a self-gravitating scalar field coupled to
matter through the matter metric ${\hat g}_{\mu\nu}$ with the
action
\begin{equation}
S=S_{\rm grav}+S_{\phi}+\hat{S}_{\rm M},
\end{equation}
where
\begin{equation}
S_{\rm grav}=-\frac{1}{\kappa}\int d\mu (R[g]+2\Lambda).
\end{equation}
Here, $\kappa=16\pi G/c^4$, $\Lambda$ is the cosmological constant,
$d\mu=\sqrt{-g}d^4x$, $\mu=\sqrt{-g}$, $\hat\mu=\sqrt{-{\hat g}}$ and
$c$ denotes the currently measured speed of light.
We have the relation $\mu=\sqrt{K}\hat{\mu}$ where
$K=1-B\hat{g}^{\mu\nu}\partial_\mu\phi\partial_\nu\phi$. We use the metric signature
$\eta_{\mu\nu}={\rm diag}(+1,-1,-1,-1)$ where $\eta_{\mu\nu}$ is the flat
Minkowski metric.

The minimally-coupled scalar field action is
\begin{equation}
S_{\rm \phi}=\frac{1}{\kappa}\int d\mu\,
\Bigl[\frac{1}{2}g^{\mu\nu}\partial_\mu\phi\partial_\nu\phi-V(\phi)\Bigr].
\end{equation}
From this action we can derive
\begin{equation}
\bar{g}^{\mu\nu}\hat{\nabla}_\mu\hat{\nabla}_\nu\phi+KV^\prime
[\phi]=0,
\end{equation}
where
\begin{equation}
{\bar g}^{\mu\nu}={\hat
g}^{\mu\nu}+\frac{B}{K}{\hat\nabla}^\mu\phi{\hat\nabla}^\nu\phi-\kappa
B\sqrt{K}{\hat T}^{\mu\nu},
\end{equation}
and ${\hat\nabla}^\mu$ denotes the covariant derivative with
repsect to ${\hat g}_{\mu\nu}$.

\section{\bf Bimetric Gravity Cosmology}

We can choose either ${\hat g}_{\mu\nu}$ or $g_{\mu\nu}$ to be
comoving metric frames in an FRW universe. If we choose ${\hat g}_{\mu\nu}$
as the comoving metric, then
\begin{equation}
d{\hat s}^2\equiv {\hat g}_{\mu\nu}dx^\mu dx^\nu
$$ $$
=c^2dt^2-R^2(t)\biggl(\frac{dr^2}{1-kr^2}+
r^2(d\theta^2+\sin^2\theta d\phi^2)\biggr),
\end{equation}
and
\begin{equation}
\label{gravitonspeed}
ds^2\equiv g_{\mu\nu}dx^\mu dx^\nu
$$ $$
=c^2(1-\frac{B}{c^2}\dot\phi^2)dt^2
-R^2(t)\biggl(\frac{dr^2}{1-kr^2}+
r^2(d\theta^2+\sin^2\theta d\phi^2)\biggr).
\end{equation}
From (\ref{gravitonspeed}) the speed of
gravitational waves (gravitons) is
given by $c_g(t)=c(1-\frac{B}{c^2}\dot\phi^2)^{1/2}$, while
the speed of light is constant. Alternatively, if we choose
$g_{\mu\nu}$ as the comoving metric, we have
\begin{equation}
ds^2\equiv g_{\mu\nu}dx^\mu dx^\nu
$$ $$
=c_g^2dt^2
-R^2(t)\biggl(\frac{dr^2}{1-kr^2}+r^2(d\theta^2+\sin^2\theta
d\phi^2\biggr)
\end{equation}
and
\begin{equation}
\label{VSLeq}
d{\hat s}^2\equiv {\hat g}_{\mu\nu}dx^\mu dx^\nu
$$ $$
=c^2(1+\frac{B}{c^2}\dot\phi^2)dt^2
-R^2(t)\biggl(\frac{dr^2}{1-kr^2}+
r^2(d\theta^2+\sin^2\theta d\phi^2)\biggr).
\end{equation}
Now, the speed of light is given by the time dependent quantity
$c_\gamma(t)= c(1+\frac{B}{c^2}\dot\phi^2)^{1/2}$, while the speed of
gravitational waves $c_g=c$ is constant.

In general relativity (GR), one can always
perform a diffeomorphism transformation which removes the time dependence of the
speed of light. In the bimetric gravity theory there are two speeds: one associated
with the speed of light (photons) $c_\gamma$ in the VSL metric frame, and
another with the speed of gravitational waves (gravitons) $c_g$ in the
gravitational metric frame with the dimensionless ratio
$\gamma=c_g/c_\gamma$. In contrast to GR, we cannot simultaneously remove
the time dependence in both $c_\gamma$ and $c_g$ by a diffeomorphism
transformation. Thus, by regauging clocks, we cannot make both $c_\gamma$
and $c_g$ simultaneously constant. This makes the time dependence of
either $c_\gamma$ or $c_g$ a non-trivial feature of the theory.

We will first consider the metric frame in which the
matter metric ${\hat g}_{\mu\nu}$ is comoving. Then, the Friedmann equation
is of the form
\begin{equation}
\label{Friedmann}
H^2+\frac{c_g^2k}{R^2}=\frac{8\pi{\bar G}}{3}\rho_M
+\frac{1}{3}c_g^2{\bar\Lambda}.
\end{equation}
Here, the effective gravitational constant ${\bar
G}(t)=G(1-\frac{B}{c^2}\dot\phi^2)^{3/2}$ and the effective
cosmological constant is
\begin{equation}
{\bar\Lambda}=\Lambda+\frac{1}{2}\biggl(\frac{1}{2c_g^2}\dot\phi^2+V[\phi]\biggr).
\end{equation}

When the speed of gravitons $c_g(t)\rightarrow 0$ in
the early universe, then ${\bar G}(t)\rightarrow 0$ and the
Friedmann equation (\ref{Friedmann}) becomes \begin{equation}
H^2=\frac{1}{12}\dot\phi^2, \end{equation} where we have set
$\Lambda=V[\phi]=0$. We can show by solving the wave equation
for $\phi$ for a radiation universe with the equation of state,
$p=c^2\rho/3$, that $\dot\phi^2\rightarrow {\rm constant}$ as
$t\rightarrow 0$, so that we obtain an initially inflationary
solution $R\sim \exp(At)$ where $A={\dot\phi}/\sqrt{12}={\rm
constant}$. Note that the derivation of initial inflation {\it
does not depend significantly} on a choice of potential
$V[\phi]$. The initial inflationary phase is produced by the
assumption that the speed of gravitons becomes very small in the
early universe.

We shall now choose the gravitational metric $g_{\mu\nu}$ as the comoving
frame. In the ${\hat g}_{\mu\nu}$ matter metric frame (we call it the VSL
metric frame) we have:
$K(t)=1/(1+\frac{B}{c^2}\dot\phi^2)$, so that
$c_\gamma(t)=cK^{-1/2}(t)$ and the speed of photons
$c_\gamma(t)\rightarrow \infty$ as $K(t)\rightarrow 0$.

The matter stress-energy tensor
(using $\hat{u}^0=c$) is
\begin{equation}
\hat{T}^{00}=K\rho,\quad  \hat{T}^{ij}=\frac{p}{R^2}\gamma^{ij},
\end{equation}
and the conservation laws give
\begin{equation}\label{eq:cosm cons}
\dot{\rho}+3H\Bigl(\rho+\frac{p}{c^2}\Bigr)=0.
\end{equation}
We adopt the radiation equation of state $p=\frac{1}{3}c^2\rho$.
It is useful at this point to introduce the following quantities
derived from the constant $B$: $H_B^2=\frac{c^2}{12B}$ and
$\rho_B=1/(2\kappa c^2 B)$, where the latter comes from
$H_B^2=(1/6)\kappa c^4\rho_B$

The Friedmann equation is given by
\begin{equation}
\label{Friedeqn}
 H^2 +\frac{kc^2}{R^2} =
 \frac{1}{3}c^2\Lambda
+\frac{1}{6}\left(\frac{1}{2}\dot{\phi}^2+c^2V[\phi]\right)
 +\frac{1}{6}\kappa c^4 \sqrt{K}\rho.
\end{equation}
The scalar field equation is
\begin{equation}
\label{eq:biscalar eqn}
 (1-\kappa c^2 B K^{3/2}\rho)\ddot{\phi}
 +3H( 1+ \kappa B \sqrt{K}p)\dot{\phi}
 +c^2 V^\prime[\phi]=0.
\end{equation}

We postulate that in the very early universe
$K$ is very small, so that $B\dot{\phi}^2\gg c^2$ and
$c_\gamma(t)\rightarrow\infty$. We will set $V[\phi]=0$ and $\Lambda=0$.
This gives as $K\rightarrow 0$ the Friedmann equation
\begin{equation}
H^2+\frac{kc^2}{R^2}=\frac{1}{12}{\dot\phi}^2
\end{equation}
and
\begin{equation}
\label{scalareq}
{\ddot\phi}+3H{\dot\phi}=0.
\end{equation}

Eq.(\ref{scalareq}) has the solution
$\dot{\phi}=\sqrt{12}H_B\left(\frac{R_{\mathit{pt}}}{R}\right)^3$,
where we have chosen the arbitrary constant of integration
$R_{\mathit{pt}}$ to parameterize the time at which $K=1/2$.
This gives
$K=1/[1+(R_{\mathit{pt}}/R)^6]$,
so that it indicates the time at which the
gravitational and matter metrics are close to coinciding. The
subscript $\mathit{pt}$ indicates that this is the `phase
transition', where standard local Lorentz symmetry with a single
light cone will  be restored--it is also the end of the period
that will appear as inflation in the comoving gravitational
metric frame.

The Friedmann equation is then
\begin{equation}
\label{reducedFriedeqn}
H^2 = H_B^2\left(\frac{R_{\mathit{pt}}}{R}\right)^6
\end{equation}
and the biscalar field $\phi$ dominates at early times
yielding $R\sim t^{1/3}$ and $H\sim 1/3t$.

This is not an inflationary solution, and so would
seem to not solve the horizon and flatness problems. We are working in the
comoving gravitational metric frame and the speed of light propagation is
much larger than the speed of gravitational waves $c_g=c$. The
coordinate distance that light would travel since the initial singularity
is
\begin{equation}
d_H=c\int\frac{dt}{\sqrt{K}R}\propto \frac{1}{t^{1/3}},
\end{equation}
where we used $R\sim t^{1/3}$. This diverges as $t\rightarrow 0$, showing
that there is no matter particle horizon in spacetime. This solves the
horizon problem and correlates fluctuations `born' superhorizon. The
flatness problem can also be solved, because $\Omega_\phi\rightarrow 1$ as
$t \rightarrow 0$ (see ref.~\cite{Moffat}) where
$\Omega_\phi=\dot\phi^2/12H^2$.

\section{\bf Derivation of Primordial Scalar Power Spectrum}

In the comoving gravitational metric $g_{\mu\nu}$, the
significant decrease in the radiation density $\rho_r$ due to $K\rightarrow
0$ as $t\rightarrow 0$ gives a minimally-coupled Einstein-Klein-Gordon
field, and so scalar mode fluctuations $\phi=\phi_0(t)+\delta\phi(t,{\vec
x})$ about the background cosmological solution can be determined. Ignoring
gravitational back-reaction, the
equation for the perturbation of the biscalar field is
\begin{equation}
\label{eq:phi pert}
\frac{d^2\delta\phi_{\vec{k}}}{dt^2} +3H\frac{d
\delta\phi_{\vec{k}}}{dt}
+\frac{c^2\vec{k}^2}{R^2}\delta\phi_{\vec{k}}=0,
\end{equation}
where $\delta\phi({\vec x},t) = (2\pi)^{-2/3}\int
d^3k\,\exp[-i\vec{k}\cdot\vec{x}]\delta\phi_{\vec{k}}$. The
solution for our spacetime is given by Bessel functions
\begin{equation}
\delta\phi_{\vec{k}} =
A(\vec{k})J_0(\xi)+B(\vec{k})Y_0(\xi),\quad
\xi=\frac{ck}{R_{\mathit{pt}}H_B}\left(\frac{R}{R_{\mathit{pt}}}\right)^2.
\end{equation}

Since $\xi > 1$ corresponds to the modes passing back inside
the horizon at the present time, we see that in the early
universe modes of interest satisfy $\xi\ll 1$. Thus, we cannot assume a
scenario where the modes are `born' in the quantum vacuum in the early
universe. {\it Instead, we assume that the fluctuation modes are born
`superhorizon' in a ground state}.

Hollands and Wald~\cite{Hollands} assume that modes are
`born' or `emerge' from a fundamental description of spacetime, rather
than prior to some time at scales smaller than a length scale $\ell_0$.
When the modes pass through the constant sound horizon:
$c_\gamma(t)/H(t)\sim\sqrt{B}\dot\phi/H(t)\sim {\rm const.}$
($c_\gamma(t)\sim 1/t$ and $H(t)\sim 1/t$). The modes are described by  a
normalized plane wave of the flat spacetime wave operator
\begin{equation}
\label{classicalwave}
\delta\phi_{\vec{k}}(t_k) = \sqrt{\frac{\kappa \hbar c^2}{(2\pi R_k)^32\omega_k}}
\cos(\omega_k t_k -\vec{k}\cdot\vec{x} +\delta),\quad
\omega_k =\frac{ck}{R_k},
\end{equation}
where $R_k=R(t_k)$ is the scale at which a mode is born.

The fluctuations born superhorizon {\it are causally correlated
by the large speed of light or superluminal VSL mechanism in the early
universe.}

If one assumes that one should use this as initial data, then we
match not only the initial perturbation but also its time derivative.
Doing so and keeping only the dominant contribution as $\xi\rightarrow 0$
gives
\begin{equation}
\delta\phi_{\vec{k}} \approx
\sqrt{\frac{9\kappa \hbar c^2}{(2\pi R_k)^3 32\omega_k}}
\cos(\omega_k t_k -\vec{k}\cdot\vec{x} +\delta)
\ln(\xi_k)
J_0(\xi),
\end{equation}
where $\xi_k$ represents $\xi$ evaluated at $R=R_k$.
The new contribution here results from the fact that when matching onto the initial
state and its derivative the Bessel function $Y_0(z)$ is logarithmically divergent
when $\xi\rightarrow 0$. The Hollands and Wald wave
function did not have this additional logarithmic term, which will lead to slight
deviations from a scale invariant spectrum.

We obtain the spectrum of scalar field fluctuations to be
${\cal P}_{\delta\phi}
=\frac{9}{2(2\pi)^3}\break\left(\frac{\ell_P}{\ell_0}\right)^2
\ln^2(\xi_k)$, and the curvature power spectrum is then found in the usual
way (recall that $\dot{\phi}^2=12 H^2$ in this spacetime):
\begin{equation}
{\cal P}_{\cal R}
=\frac{H^2}{\dot{\phi}^2}{\cal P}_{\delta\phi}
=\frac{3}{8(2\pi)^3}\left(\frac{\ell_P}{\ell_0}\right)^2
\ln^2(\xi_k),
\end{equation}
where we have used the Hollands-Wald condition $R_k=k\ell_0$ and
$\xi_k=\sqrt{12B}\break\ell_0^2k^3/2R_{pt}^3$.

From this the spectral index is calculated to be
$n_s=1+\frac{d\ln \mathcal{P}_{\mathcal{R}}}{d\ln k}
=1+\frac{6}{\ln(\xi_k)}$
and the running of the spectral index is calculated from:
$\alpha_s=dn_s/d\ln k$, from which we find the relation
$\alpha_s=-\frac{1}{2}(1-n_s)^2$.

In the large scale limit, we therefore have
\begin{equation}
\delta_H= \frac{2}{5}\sqrt{\mathcal{P}_{\mathcal{R}}}
\approx
\frac{2}{5}\sqrt{\frac{3}{8(2\pi)^3}}\biggl(\frac{\ell_P}{\ell_0}\biggr)
\vert\ln(\xi_k)\vert.
\end{equation}

\section{\bf Comparison with the Data}

Assuming that $\ell_0\approx \sqrt{12B}$ and evaluating the
logarithm at the pivot point, $c k \sim 7 R_0H_0$, this simplifies to
$\delta_H\approx \ell_P/\ell_0 \approx 10^{-5}$,
and so we can match the amplitude of the CMB fluctuations by assuming that
$\ell_0\sim \sqrt{12B} \approx 10^5 \, \ell_P$.
This choice predicts~\cite{Clayton} $n_s \approx 0.97,\quad
\alpha_s \approx -4.5 \times 10^{-4}$.
These results agree well with the WMAP data $n_s=0.99\pm
0.04$~\cite{Spergel}.

We can predict the acoustical waves power spectrum and fit
it to the corresponding WMAP data that can be detected at the surface of
last scattering. By adopting the $\Lambda$CDM data from WMAP and
our prediction $n_s=0.97$, we obtain a fit
to the data as good as the fit obtained from an inflationary
model using an appropriate slow-roll potential for the inflaton.

An important piece of data obtained from the WMAP observations is
the cross polarization E-T temperature power. Since our primordial
fluctuations are born `superhorizon' and are causally correlated by the
superluminal VSL mechanism early in the universe, we obtain a good fit to
the cross correlated polarization data. Again
the fit to the data is competetive with inflationary model
predictions. This data confirms that a superluminal `causal'
mechanism, such as that provided by a generic inflationary model
or the VSL bimetric gravity model, is necessary to fit the E-T
polarization data obtained by WMAP.

\section{\bf Conclusions}

We have shown that the VSL mechanism in the bimetric gravity
model, choosing the gravitational metric as comoving, predicts a scale
invariant power spectrum with a spectral index $n_s=0.97$, which agrees
well with the WMAP data. The mechanism for producing this result is quite
different from standard inflationary models, in which a `slow rolling'
potential is chosen from a large possible number of potential
models~\cite{Linde}.

The basic postulate is that $c_\gamma(t)$ becomes very large in
the early universe, solving the flatness and horizon problems. It is
possible that the VSL mechanism can avoid the initial fine-tuning problems
of generic inflationary models.

A possible observational test to distinguish the bimetric gravity
model from standard slow-roll inflationary models is to observe the tensor
mode (B-mode) and the ratio $r=T/S$ in the power spectrum. The predictions
$r$ in the VSL bimetric gravity theory and generic inflation models may be
sufficiently different in magnitude to decide which model agrees better
with observations. However, the pure gravitational tensor modes have not
been detected by WMAP.

In the VSL metric frame in the bimetric gravity model, we are not required
to have a large vacuum energy in the initial phase of the universe, so
that no fine tuning is needed to obtain the tiny cosmological constant
that is favored by the WMAP and supernovae data. This is in contrast to
inflationary models which require a very large vacuum energy in the early
universe to generate enough e-folds of inflation.
\vskip 0.2 true in
{\bf Acknowledgments}
\vskip 0.2 true in
This research was supported by the Natural Sciences and Engineering
Research Council of Canada.
\vskip 0.2 true in
 \end{document}